\documentclass[preprint,12pt]{article}
\usepackage[utf8]{inputenc}
\usepackage{authblk}
\usepackage{amsmath,amssymb}
\usepackage{graphicx}
\usepackage{xcolor}

\title{Plasma free bubble cavitation in water by a 2.9 µm laser for bioprinting applications}
\author[1]{Shubho Mohajan}
\author[1]{Jean-Christophe Delagnes}
\author[1]{Baptiste Allisy}
\author[2]{Antonio Iazzolino}
\author[2]{Bertrand Viellerobe}
\author[,1]{Stéphane Petit\thanks{Corresponding author: \texttt{stephane.petit@u-bordeaux.fr}}}

\affil[1]{\textit{CELIA Centre Lasers Intenses et Applications UMR 5107 Université Bordeaux-CNRS-CEA 33400 Talence, France}}
\affil[2]{\textit{POIETIS, Bioparc Bordeaux Métropole, 27, allée Charles Darwin 33600 Pessac, France}}

\begin{document}

\maketitle

\textit{The following article has been submitted to Applied Physics Letters. \\ After it is published, it will be found at https://aip.scitation.org/journal/apl}

\begin{abstract}
We investigate the dynamics of the cavitation bubble induced by 2.9 µm mid-IR laser pulses (10 ns, 10-50 µJ) resulting in a plasma-free direct fast heating of water due to a strong vibrational absorption. We establish a direct correlation between the laser fluence (up to $6\text{ J.cm}^{-2}$) and the maximum bubble radius (up to 200 µm). From experimental data, key parameters (threshold energy, internal pressure) can be retrieved by simulations including the water absorption saturation at 2.9 µm. At a fluence of $6\text{ J.cm}^{-2}$, we obtain 13 \% of the laser energy converted to a bubble energy and we can predict that operating at higher fluence $>10\text{ J.cm}^{-2}$ will lead to a maximum  of 20 \% conversion efficiency. This results open the door to bioprinting applications using direct absorption of the laser radiation without any additional absorber.
\end{abstract}

\newpage

Bioprinting of living cells has become in the last decade a new key technology to create complex 3D structures of bio-compounds \cite{Guillemot2010} applied  to regenerative medecine of skin, bones \cite{keriquel_situ_2017}, vascularized tissues \cite{zhang_recent_2021} or possibly artificial organs for transplantation \cite{Li2016}.

Laser Assisted Bioprinting (LAB) is a beautiful biophotonic application involving laser technology, light-matter interaction, fluid mechanics and biology \cite{Li2021}. LAB consists in transferring the laser energy to an aqueous thin-film containing living cells to be printed. The deposited energy triggers a cavitation bubble causing the fluid to form a jet at the free surface and carrying the cells. The jet evolution solely depends on fluid dynamics but it finds its origin in the laser induced cavitation bubble. Stable jets in homogeneous liquids are usually formed when the maximum cavitation bubble radius is close from the free surface \cite{patrascioiu_laser-generated_2014}. 

A widespread method of energy transfer consists in using a substrate coated with a thin layer of absorbing material supporting a 100-200 µm thick film of bio-ink. Various materials have been used as a "sacrificial" absorber such as metals, metal-oxides or polymers \cite{Li2021}. The electrons are highly excited by the laser absorption leading to a plasma generation. The energy transferred to the fluid thus forms a bubble initiating the jet. However, printed tissues require complex patterns drawn by a scanner. Both the refreshing of the ablated sacrificial layer and the replacement of the bio-ink after few tens of patterns are the current limitation of the standard LAB approaches.

To circumvent this limitation, the plasma can be created in the medium itself. When near IR 10 µJ femtosecond \cite{duocastella_film-free_2010} and picosecond \cite{petit_femtosecond_2017} pulses are focused underneath the liquid surface at an intensity of $10^{11}\text{-}10^{13} \text{ W/cm}^{2}$, multiphoton or tunnel ionisation creates free electrons. This results in an optical breakdown. Its dynamics has revealed that the bubble formation \cite{vogel_energy_1999, vogel_mechanisms_2005} originates from the fast heating caused by the electron-ion recombination \cite{schaffer_dynamics_2002}. Yet, spatial heterogeneity of media such as bio-ink combined with the highly nonlinear nature of the process hampers the repeatability of energy deposition via plasma formation.

A direct method for heating of liquids is the 'one photon' absorption by water in some spectral bands. 
Depending on the physical and chemical processes following absorption, the survival of living cell can be affected by critical parameters
such as pressure, temperature, or chemical composition.

Although fast and highly localized, water photolysis and the risk of DNA-degradation limits in practice the use of UV-lasers \cite{zhang_effects_2017}. In the visible and near IR, liquids with adjuvent as absorber support the generation of sub-mm diameter bubbles by thermocavitation \cite{rastopov_cluster_1990,ramirez-san-juan_time-resolved_2010,padilla-martinez_optic_2014}. Superheated water (typ. up to 200-300 °C) forms a metastable liquid evolving in its phase diagram following a spinodal curve up to the critical point. This leads to an explosive liquid evaporation \cite{padilla-martinez_optic_2014,vogel_femtosecond-laser-induced_2008, xiong_bubble_2015} causing the bubble to appear.

Taking advantage of the strong absorption of water in the mid-IR range allows adjuvent- and plasma-free energy deposition. We thus investigate laser-induced bubbles in the $\sim$2.9 µm vibrational absorption band of water. IR generated bubbles have been firstly observed with a 1 J-50 ns-10.6 µm  CO$_2$ laser \cite{emmony_interaction_1976}. The $\sim$2.9 µm band has been exploited to strongly ablate liquid surface by  10 mJ-70 ns-2.97 µm Er:YAG  with fluence up to a 5.6 $\text{J/cm}^{2}$ for laser-biological tissue applications \cite{apitz_material_2005}. Due to the ultrafast thermalization of the O-H stretch \cite{Fran2010}, picosecond pulses lower the required fluence down to 0.75 $\text{J/cm}^{2}$ with 1 mJ pulse energy \cite{franjic_laser_2009}.

More recently, 45 ns-2.85 µm millijoule-class laser have been used to study shock waves and bubble cavitation \cite{pushkin_cavitation_2018} for fluence ranging from 0.45 to 2 $\text{J/cm}^{2}$. In such conditions, bubbles with maximum radius from 0.4 to 1.2 mm have been reported but their large size is not compatible with bioprinting applications. However, huge progress in the mid-IR laser technology make nowadays available robust diode-pumped industrial lasers sources based on solid-state (Cr:ZnSe), parametric oscillators or mid-IR Er-doped fibers delivering 1 to 100 µJ  energies. Although promising to produce small bubbles ($<$ 200 µm radius) key for bioprinting applications, this range of low energy mid-IR pulses is -- to the best of our knowledge -- still largely unexplored and is the scope of the present letter. 

\begin{figure}[htbp]
\centering
\includegraphics[width=1\textwidth]{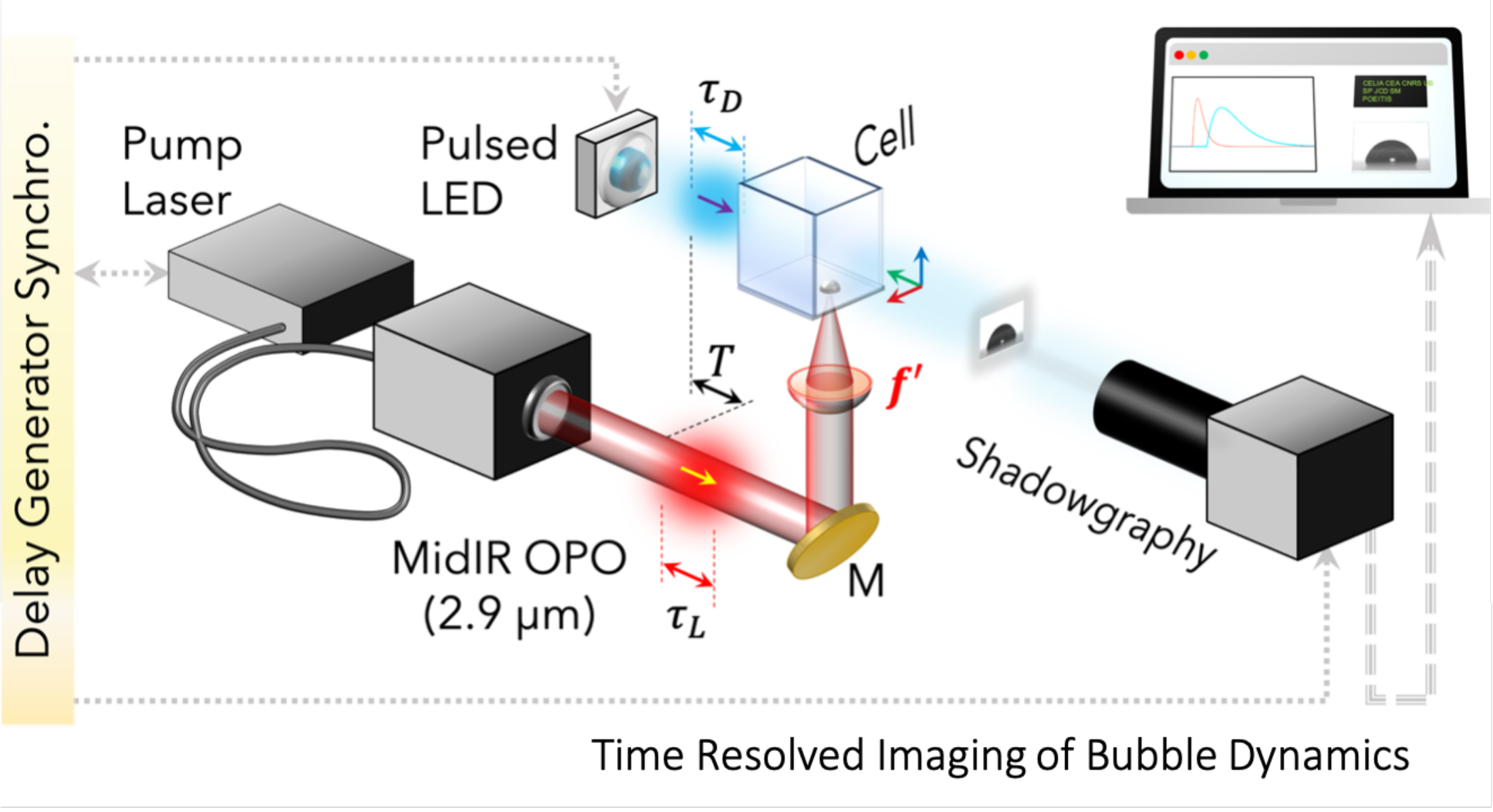}
\caption{\label{Fig1_ArXiV} Experimental set-up: a laser emitting 10 ns pulses at 2.9 µm is focused in a cell filled with water. The cavitation bubble are imaged by shadowgraphy on a CMOS camera and the dynamics are resolved by a Time Resolved Imaging system synchronized with the laser. Bubble radii are then extracted by automatic image processing. }
\end{figure}

The experimental setup is depicted in Fig.\ref{Fig1_ArXiV}. The mid-IR source consists of an Optical Parametric Oscillator (OPO) (APE-GmbH) pumped by a 10 ns-1064 nm diode-pumped Nd:YAG Q-Switched laser (CANLAS Laser Processing GmbH). The laser operates from single-shot up to 1 kHz with a maximum pulse energy of 55 µJ at 2.9 µm. The experiments have been performed at few Hz to ensure that neither cumulative thermal effect nor residual acoustic wave from previous pulse disturb the liquid. The output energy is ajusted by a build-in variable attenuator down to 0.5 µJ with a shot-to-shot energy stability of 2.5 \% rms. A five-wall polished spectroscopic quartz cell is filled with de-ionized water. The laser beam is focused upwards on the bottom of the cell using AR-coated CaF$_2$ lenses (20, 40 or 50 mm focal length) or ZnSe microscope objective (12 mm focal length). From the output of the laser and the cell-water interface including the water-atmospheric absorption ($\alpha_{\textrm{atm}}=2.94\times 10^{-3}\text{ cm}^{-1}$), the overall transmission is 77 \%.

A time resolved imaging is used to acquire sequences of images in a shadowgraphic configuration. It consists of a collimated-500 ns (FWHM)-pulsed blue LED propagating through the spectroscopic cell's sidewalls and a triggered CMOS camera (UI-3260CP, IDS). A delay generator (SRS DG535) is used to externally trigger the laser and to synchronize the delays between the laser shot, the pulsed blue diode and the camera. A homemade program manages the delays, the camera and the recording of image sequences. The images are calibrated with a resolution test target (USAF) and a zoom lens assembly (OPTO) allowing a 3.7 maximum magnification resulting in a resolution of 3.57 µm/pixel.

For each laser energy, the bubble movie is reconstructed by recording 50 images at each delay $t$ after the laser pulse thus providing enough statistics for averaging shot-to-shot fluctuations. Each image was processed by Fuji \cite{schindelin_fiji_2012} and the long and short axis radii (typ. 6 \% r.m.s. accuracy) are extracted to track the bubble rim dynamics.

\begin{figure}[htbp]
\centering
\includegraphics[width=1\textwidth]{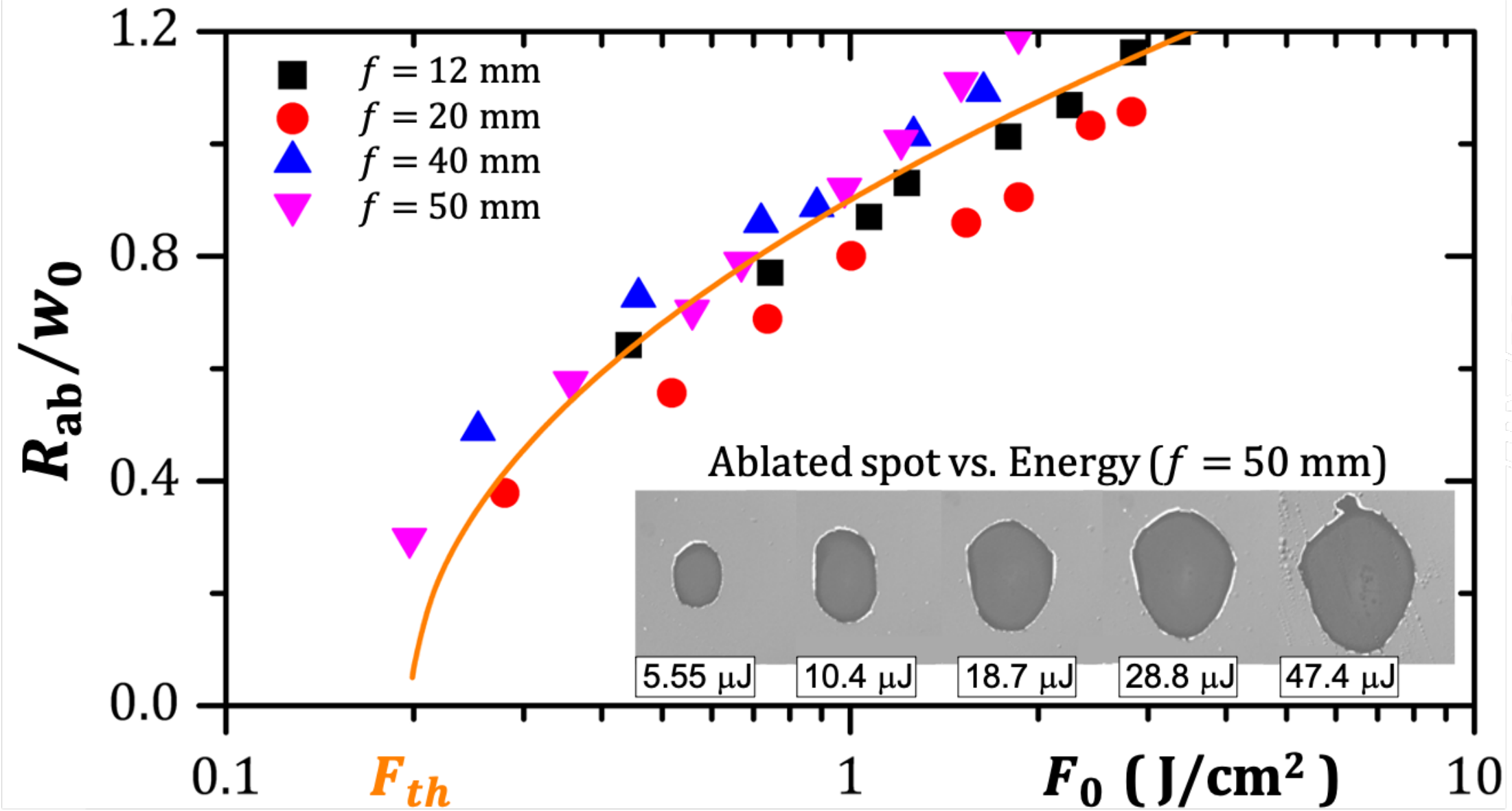}
\caption{\label{Fig2_ArXiV} Average ablated radius $R_\textrm{ab}=\langle D \rangle/2$ relative to the beam waist $w_0$ vs. fluence $F_0$ for 12, 20, 40, and 50 mm lenses. Inset: Evolution of the ablation spot area on 20 nm gold layer as a function of laser energy for 50 mm lens (the corresponding energy is given for each image). }
\end{figure}

To estimate the fluence, we first determined the focal spot size. Since direct measurement of the focal spot is not feasible with standard large-pixel middle-infrared imaging sensors, we estimated the beam waist by using ablation spot size imaging. A 20 nm thick gold layer is ablated by a single shot of 2.9 µm laser pulse for different laser energies $E_0$ and different focal lengths. Then, the ablation spots are imaged using a microscope (NIKON) as shown in inset in Fig.\ref{Fig2_ArXiV}.

The measured ablation spot is plotted against laser energy (Fig.\ref{Fig2_ArXiV}). For a Gaussian beam in the vicinity of the confocal region, the fluence in polar coordinates reads $F(\rho)=F_0\cdot\exp(-2\rho^2/w_0^2)$ where $F_0=2 E_0/\pi w_0^2$. The relationship between the mean diameter $\langle D \rangle$ of the ablated area and fluence $F_0>F_{th}$ above gold ablation threshold  $F_{th}$ is fitted by the equation \cite{Liu:82}, $\langle D \rangle^2 = 2 w_0^2\cdot\ln(F_0/F_{th})$.

The extracted beam waist values $w_0$ respectively are 22.5, 25, 28, and 31.5 µm for lenses having focal length of 12, 20, 40, and 50 mm respectively, while the retrieved threshold $F_{th}= 0.2 \text{ J.cm}^{-2}$ is consistent with the value reported in the literature \cite{GoldThreshold}.  These measurements are more accurate than the estimate based on the $M^2$ values independently obtained (not shown) which underestimated the spot size. Fluence regimes ranging from 0.1 to 6 $\text{ J.cm}^{-2}$ have then been investigated.

\begin{figure*}
\includegraphics[width=1\textwidth]{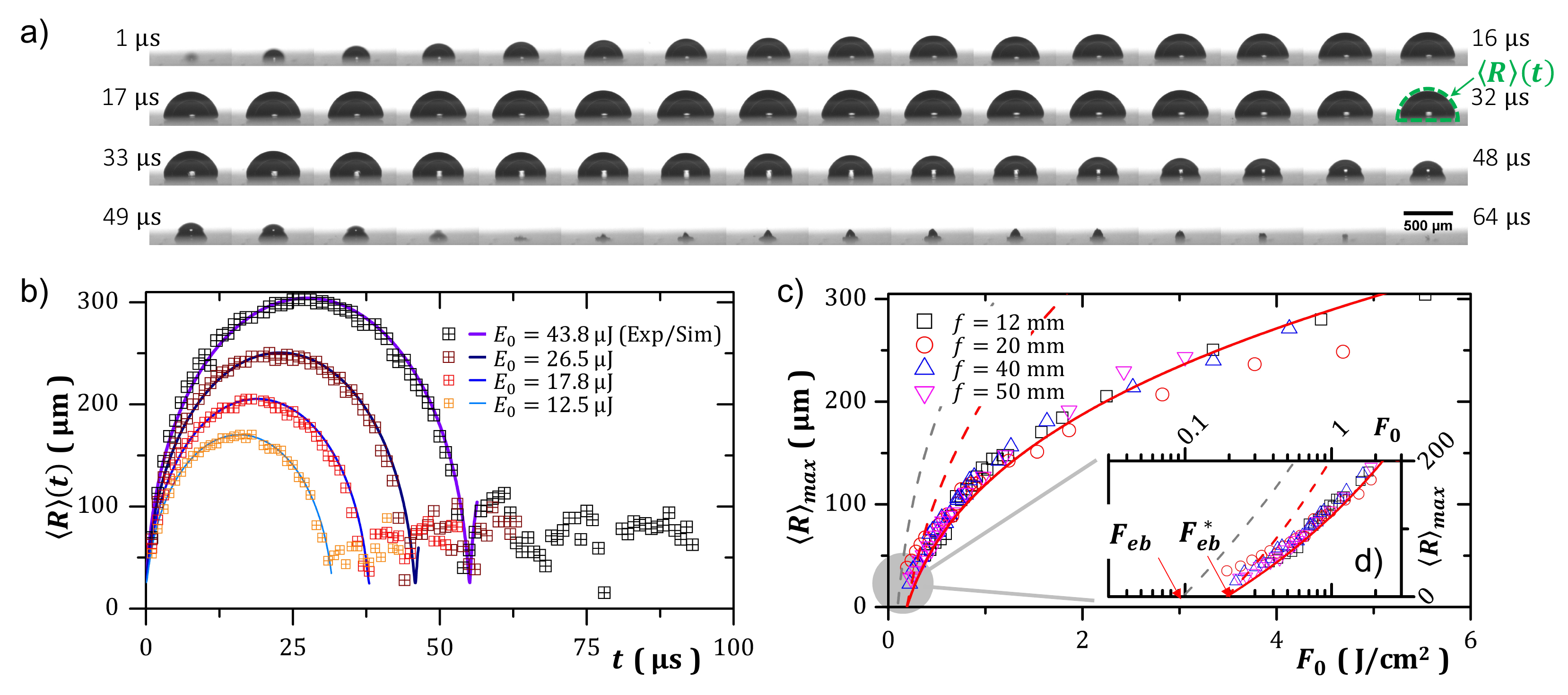}
\caption{\label{Fig3_ArXiV} (a) Time resolved images of bubble dynamics with 1 µs time step using 12 mm focal length lens, 43.8 µJ energy.(b) Experimental analysed results of bubble dynamics for 43.8 µJ, 26.5 µJ, 17.8 µJ, 12.5 µJ ($\boxplus$)
and a 12 mm focal length lens.  Line plots represent the simulations by solving Rayleigh-Plesset equation with $R_\textrm{eb}^*$ and $P_0$ as initial values. (c)  Experimentally measured maximum bubble as a function of laser fluence for the different lenses and energies (open symbols). Lines represent the expected maximum bubble radius when the conversion of the excess energy above the explosive boiling energy of water in bubble energy is 100 \% without (gray dashed line) or with (red dashed line) the saturation of absorption or 20 \% (red line) with saturation. (d) zoom in semi-log scale of (c) near threshold showing the meaningful contribution of the saturation of water absorption.}
\end{figure*}

As already reported \cite{emmony_interaction_1976,padilla-martinez_optic_2014,pushkin_cavitation_2018}, a (quasi)hemispherical bubble of radius $R$ is created and further expands. In Fig.\ref{Fig3_ArXiV}(a), the bubble is sticked on the CaF$_2$-water interface by the strong absorption of water at 3 µm ensuring a constant location of the bubble which is highly beneficial for the repeatability of the process. The bubble grows up to a maximum radius $R_{max}$ and starts collapsing beyond this maximum radius. In the vicinity of the first collapse event, Rayleigh-Taylor instability affecting the bubble shape  is also observed, thus reducing the accuracy in bubble size extraction. After this first collapse, the subsequent bouncing dynamics is not influenced by the initial excitation anymore, and follows what has already been reported in the literature \cite{zhong_model_2020}. Up to three bounces were obtained for the highest energy available (43.8 µJ). We here focus our analysis on the first expansion-climax-collapse cycle as it is governed by the initial conditions resulting from the laser energy deposition investigated here.

In Fig.\ref{Fig3_ArXiV}(b) where the averaged bubble radius $R$ as a function of time $t$ is plotted, the dynamics follows the Rayleigh-Plesset equation:
\begin{equation}\label{RP_Equation}
    \ddot{R}=-\frac{3\dot{R}^2}{2R}+\frac{1}{\rho_l R}\left[P_B-P_a-\frac{2\sigma}{R}-\frac{4\mu\dot{R}}{R}\right]
\end{equation}
where $\rho_l=1000 \text{ kg.m}^{-3}$ is the liquid phase (water) density, $\sigma=0.072 \text{ N.m}^{-1}$ the surface tension, $\mu=10^{-3} \text{ Pa.s}$ the dynamic viscosity, $P_B$ the pressure on the bubble surface, and $P_a\simeq 10^5 \text{ Pa}$ the ambient atmospheric pressure \cite{zhong_model_2020}.

The first bounce is well reproduced (solid lines in Fig.\ref{Fig3_ArXiV}(b)) by solving the Rayleigh-Plesset Eq.~(\ref{RP_Equation}) using Euler's method \cite{yasui_acoustic_2018}. The experimentally observed first collapse time for 12.5 to 43.8 µJ energies is between 30 to 55 µs. This is in good agreement with both the numerical simulation and the Rayleigh-collapse time \cite{Rayleigh} $t_c=0.915R_{max}\sqrt{\rho_l/(P_a-P_v)}$ where $P_v=$2.33 kPa is the saturated vapour pressure.

To reveal the dependence of maximum bubble radius $R_{max}$ on laser fluence $F_0$, several bubble dynamics were obtained using different laser energies and focusing conditions. Each focusing condition leads to a different $R_{max}$ vs. energy $E_0$ relationship. However, all curves merge together when the measured averaged bubble radii $\langle R\rangle_{max}$ is plotted vs. the fluence (Fig.\ref{Fig3_ArXiV}(c)). This nonlinear relation between $R_{max}$ and fluence will be discussed later.

Contrasting with what can be observed with plasma-induced cavitation bubble in water, we didn't observe any remaining bubbles in the cell even for the highest fluence investigated here. The absence of gas creation or accumulation between laser shots (repetition rate here typ. from 1 to 4 Hz) along with the absence of any recombination plasma light strongly indicates that the process is plasma-free. No residual gas products from water photolysis such as H$_2$, O$_2$, and O$_3$ that might affect the living cells have been observed. This is consistent with the fact that the maximum density of energy deposited $\alpha^*(F_0)F_0\simeq 20 \text{ kJ.cm}^{-3}$ at $F_0=10 \text{ J.cm}^{-2}$, while the first and the second O-H bond dissociation energy density of the water molecule in liquid phase are 27.4 and 23.6 $\text{ kJ.cm}^{-3}$ respectively. Hence, only the strong one-photon absorption at 3 µm in water is involved in the process of bubble generation.

Since the range of fluence ($> 1 \text{ J.cm}^{-2}$) leading to the observation of cavitation bubble sizes ($\gtrsim$ 100 µm) is relatively high, we need to take into account the saturation \cite{KresimirFranjic2010,Shori_Saturation} of water absorption at 2.9 µm. From the saturated transmission measurements reported \cite{Shori_Saturation}, we have fitted the Beer-Lambert's law $\textrm{d}F/\textrm{d}z=-\alpha^*(F)F$ by the following \textit{ad hoc} formula (\textcolor{blue}{see Supplementary material}):
\begin{equation}\label{Sat_Abs}
    \alpha^*(F)=\alpha_0+\Delta\alpha \cdot f/(1+f)
\end{equation} 
$\alpha^*(F)$ being the fluence dependent absorption coefficient (thereafter, the use of $^*$ means that the saturation of absorption is taken into account) where $\alpha_0=12787.3 \pm 18.1 \text{ cm}^{-1}$ is the unsaturated absorption coefficient, $\Delta\alpha= -10906.6 \pm 23.9 \text{ cm}^{-1}$ the asymptotic drop of absorption and $f=F/F_s$ with $F_s=0.136 \pm 0.006 \text{ J.cm}^{-2}$ the saturation fluence. Although different from conventional saturated formulae, not only Eq.(\ref{Sat_Abs}) better reproduces the reported results \cite{Shori_Saturation}, but it also allows us to derive analytical expressions leading through thermodynamic considerations to the exact formulation of the excess energy initially available after explosive boiling for bubble expansion  and then to the maximum bubble diameter that can be achieved.

The excited volume $V^*$ that contributes to bubble formation is defined as the volume for which the effective critical fluence $F_\textrm{eb}^*$ has a deposited energy density $u_\textrm{dep,eb}^*=\alpha^*(F_\textrm{eb}^*)F_\textrm{eb}^*$ that is exactly equal to the explosive boiling energy density ($H_\textrm{eb}\rho_l$) with no excess energy in both radial and longitudinal distributions. Including the saturation of absorption for a stigmatic Gaussian beam, we find $F_\textrm{eb}^*=0.21 \pm 0.02 \text{ J.cm}^{-2}$ which is about twice higher than the effective threshold in the absence of saturation ( $F_\textrm{eb}=0.1 \text{ J.cm}^{-2}$). Within a Gaussian beam cross-section, the maximum critical radius where explosive boiling may occur is then given by $R_\textrm{eb}^*=w_0\sqrt{\textrm{ln}(F_0/F_\textrm{eb}^*)/2}$ and corresponds to the off-axis distance where the fluence is sufficient to initiate a phase transition. The effective penetration depth $z_\textrm{eb}^*$ inside the critical radius is deduced which leads after integration to the expression of the excited volume $V^*$. Similarly, the excess energy available for bubble expansion is given by (\textcolor{blue}{see Supplementary material}):
\begin{equation}\label{Eexplosive}
    E_\textrm{ex}^*=\int_{\rho=0}^{R_\textrm{eb}^*}2\pi \bigg(F(\rho)-F_\textrm{eb}^*-H_\textrm{eb}\rho_l z_\textrm{eb}^*(\rho)\bigg)\rho\textrm{d}\rho   
\end{equation} and the maximum bubble diameter is retrieved  with the cavitation bubble energy $E_B$ expressed by Vogel \textit{et al.} \cite{vogel_energy_1999}.

We applied this procedure only based on thermodynamics to estimate the maximum cavitation bubble diameter reachable at a given laser fluence deposited in water. The results are shown in Fig.\ref{Fig3_ArXiV}(c) and (d) with the dashed and plain lines corresponding to different conditions. As highlighted by a closer inspection near threshold (Fig.\ref{Fig3_ArXiV}(d)), the absence of saturation of the absorption leads both to a lower expected threshold of bubble appearance $F_\textrm{eb}$ and a much higher increase of the bubble size as the fluence is growing  (gray dashed line) than experimentally observed. The saturation of water absorption pushes the threshold $F_\textrm{eb}^*$ exactly where it is experimentally observed but the maximum bubble size is still overestimated when 100 \% of the excess explosive boiling energy is converted into bubble energy (red dashed line). If 20 \% of the excess energy is converted in bubble energy, one can see that the estimated expected bubble diameters (red plain line) fit perfectly the experimental data over a full decade in fluence.

The energy $E_\textrm{ex}^*$ and the volume $V^*$ allow us to estimate the initial pressure $p^*$ (\textcolor{blue}{see Supplementary material}). In Fig.\ref{Fig4_ArXiV}(a), $p^*$ is in the range of several hundreds of MPa to few GPa when the maximum radius of the bubble grows up to 300 µm for the different focal lenses used in the experiment. $p^*$ is excessively high  and  it leads to an acoustic radiation \cite{wang_2016} and a shock-wave \cite{Vogel_shock_1996, Tomita_Shock_nitrogen,Shock_Wave_Probe} dissipating part of the deposited energy while surface tension and viscosity play a negligible role.

The relative amount of energy released in the shock-wave $\Delta\epsilon_{sw}=E_{sw}/E_B$ can be evaluated from the dimensionless parameter \cite{PhysRevE_ShockFraction} $$\Xi=\Delta P_{av}^{}P_{nc}^\Lambda(\rho_l c^2)^{1-\Lambda}/\Lambda^6$$ where $\Delta P_{av}=P_a-P_v$ is the pressure difference at $R_{max}$ and $P_{nc}$ is the noncondensable gas pressure at $R_{max}$, $c=1500 \text{ m.s}^{-1}$ is the speed of sound in water, and $\Lambda=1/\kappa$ according to the Keller-Miksis model \cite{KM_Model}. For a polytropic index $\kappa=1.4$, even a high residual pressure $P_{nc}\lesssim P_v$ already leads to $\Delta\epsilon_{sw}\gtrsim$ 40 \% of the total energy released in the shock. With a 6 ns-1064 nm laser focused in water at the same range of fluence, Vogel \textit{et al.} \cite{vogel_energy_1999} have reported a range between 35 \% and 70 \% of conversion to shock energy during the interaction. This implies that a substantial amount of the laser energy is carried by the generated shock wave during the interaction of the laser pulse with water regardless the wavelength of the laser and the process of the fast heating.

The rest of the energy is used to start the growth of the bubble with a lower initial pressure $P_0$. In order to evaluate this pressure, we solved the Rayleigh-Plesset for all dynamics with a varying initial radius $R_\textrm{eb}^*$ and a varying initial pressure $P_0$ with a numerical sensitivity of 1 \%,  $P_0$ being refined by least-square fitting the first cycle dynamics. In Fig.\ref{Fig4_ArXiV}(a), $P_0$ is in the range of few tens of MPa when a maximum bubble radius of 300 µm is reached for the different lenses. In the scope of bioprinting application, living cells present in the focal volume during the bubbling process will thus experience this quite high pressure could have an impact on their viability.
However, we can produce equivalent bubble radius with a lower initial pressure by using a longer focal length but a larger energy. The initial pressure $P_0$ is almost four times lower when using a 50 mm focal length rather than a 12 mm focal length (Fig.\ref{Fig4_ArXiV}(a)). Therefore, long focusing lens are likely to be  preferable for bioprinting applications as the living cell will be exposed to lower pressure.

An another important aspect is the efficiency of light-to-mechanical energy conversion process. The bubble initially grows with a constant velocity and the corresponding maximum kinetic energy $E_k=\pi\rho_l R^3 \dot{R}^2$  is calculated by fitting the initial dynamics of the bubble for each conditions. As it can be seen in  Fig.\ref{Fig4_ArXiV}(b), $E_k$ and $E_B$ are largely equal over a large range of fluence showing that most of the mechanical energy is converted in bubble energy with no losses due to friction, vorticity, nor bubble wall motion at the triple contact point between gas, liquid, and substrate.

Moreover, at low fluence, most of the laser  energy is used to heat the liquid, only a small fraction is being converted into explosive boiling. At $2 \ \mathrm{J/cm^2}$, we achieve with the 2.9 µm-10 ns laser 7 \% efficiency of bubble generation analogous to what Pushkin \textit{et al.} \cite{pushkin_cavitation_2018} have reported  with a 5 \% efficiency by using a 2.85 µm-45 ns laser at the same fluence. At $6 \ \mathrm{J/cm^2}$, we achieve 13 \% but at larger fluence close to $10 \ \mathrm{J/cm^2}$, the excess energy converges to the laser energy and in such condition a maximum of 20 \% efficiency can be expected. Operating at larger fluence than $10 \ \mathrm{J/cm^2}$ would increase the efficiency to the limit but with much bigger bubbles that will not be compatible with the bioprinting application.  

In this letter, we have investigated the dynamics of cavitation bubble in water generated by microjoule nanosecond pulses emitted at 2.9 µm with different focusing and energy conditions. As far as bioprinting applications are concerned, few directions can be drawn from this work. It clearly shows that the cavitation bubble dynamics is only governed by the deposited laser fluence in the medium. At this wavelength, only the strong single photon absorption of water is involved through the explosive boiling in the bubble generation, while no evidence for any plasma creation has been observed. The constant location of the bubble without any added artificial absorber and other chemical compounds created during the interaction is a highly beneficial advantage compared to the other methods, insuring repeatability, stability and purity. In the regime of fluence $>1 \ \mathrm{J/cm^2}$, the saturation of the water absorption must be taken into account in order to estimate correctly the volume of interaction and the energy transferred from the laser to the  bubble. In these conditions, we have shown that at high fluence ($>10 \ \mathrm{J/cm^2}$) a maximum of 20 $\%$ of the laser energy can be expected to be transferred into the bubble, the rest being converted in shock wave. For the targeted bioprinting application, longer focal length thus seem preferable in order to decrease the internal pressure in the bubble that might be experienced by the cells.

These experiments and analysis pave the way to complex structuring of living tissues by using mid-IR laser technology without any assistance of artificial absorber. It also opens the way towards a full automation of the bioprinting where the reproductibility and stability of the cell deposition in a shorter time ensure a reliable, cost-effective process compliant with the GMP grade requirements.

\section*{Supplementary Material}
See \textcolor{blue}{supplementary material} for the analytical calculations of the effective threshold for explosive boiling $F_\textrm{eb}^*$  and the excited volume  $V^*$ leading to the determination of the excess energy for explosive boiling $E_\textrm{ex}^*$  and the initial pressure $p^*$. 

\section*{Acknowledgments}
We wish to acknowledge Pierre Hericourt for the development under Python of an automatic data acquisition system managing thousands of frames, images and delays allowing fast high quality data collection and enabling reliable statistic operations. Technical support from Edlef Büttner (APE GmbH) for laser installation and  troubleshooting solving has also been much appreciated. This work has been supported by Eureka - Eurostars MIR-LAB n° E111513

\section*{Data Availability Statement}
The data that support the findings of this study are available from the corresponding author upon reasonable request.

\bibliographystyle{unsrt}
\bibliography{main}

\newpage

\begin{figure*}
\includegraphics[width=1\textwidth]{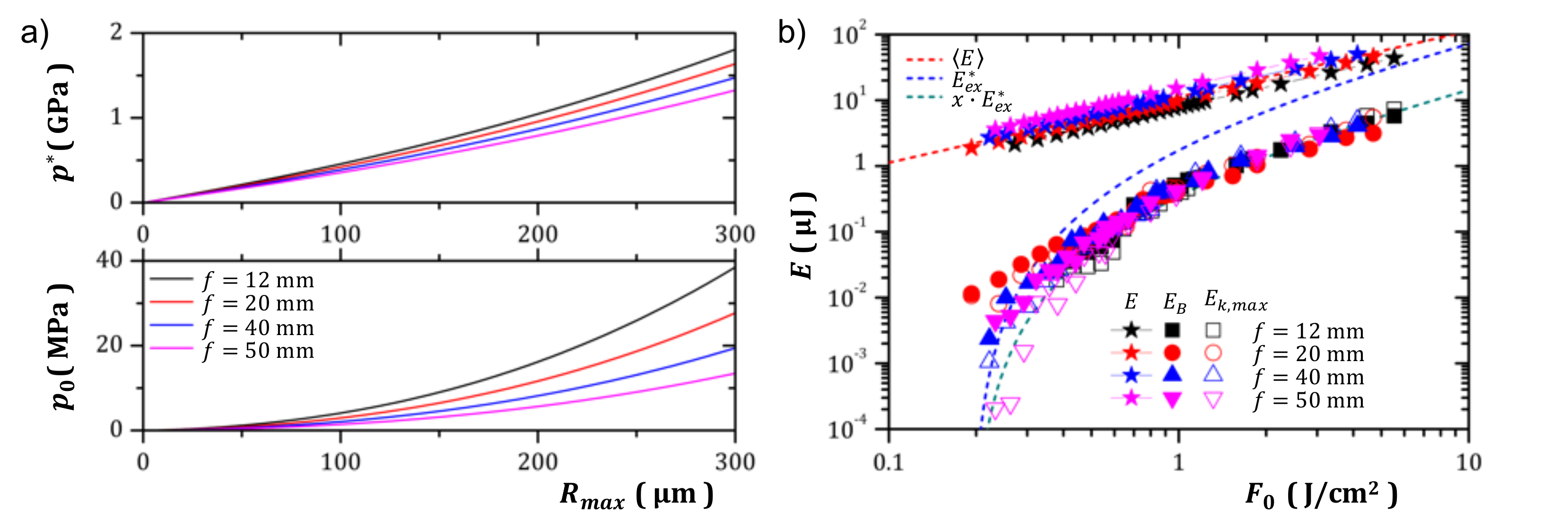}
\caption{\label{Fig4_ArXiV} (a) (Upper) Initial pressure after deposition of $E^*_{ex}$ in the interaction volume $V^*$. (Lower) Initial pressure inside the bubble calculated from Rayleight-Plecett simulation dynamics with 20 \% of the excess energy above explosive boiling energy converted in bubble energy. All pressures have been caclulated for the different lenses. (b) Bubble energy $E_B$ (filled symbols), maximum kinetic energy $E_{k,max}$ (open symbols) and total energy $E$ (filled stars) as a function of laser fluence for different lenses. Note that $E_B$ and $E_k$ are not maximum at the same time in the dynamics. The expected bubble energy with 100 \% (blue dash line) and 20 \% (green dash line) of the excess energy converted are also compared with the total energy (red dash line). For low fluence ($<$1 $\mathrm{ J/cm^2}$) most of the laser energy is dissipated in heat without explosive boiling of water. For fluences $>$10 $\mathrm{ J/cm^2}$ a maximum of 20 \% of the laser energy can be expected to be converted in bubble energy, the rest being dissipated in shock wave.}
\end{figure*}

\newpage

%%%%%%%%%%%%%%%%%%%%%%%%%%%%%%%%%%%%%%%%%%%%%%%%%%%%%%%%
%%%%%%%%%%%%%%%%%%%%%%%%%%%%%%%%%%%%%%%%%%%%%%%%%%%%%%%%
%%%%%%%%%%%%%%%%%%%%%%%%%%%%%%%%%%%%%%%%%%%%%%%%%%%%%%%%
%%%%%%%%%%%%%%%%%%%%%%%%%%%%%%%%%%%%%%%%%%%%%%%%%%%%%%%%
%%%%%%%%%%%%%%%%%%%%%%%%%%%%%%%%%%%%%%%%%%%%%%%%%%%%%%%%

\section*{Supplementary Material: Plasma free bubble cavitation in water by a 2.9 µm laser for bioprinting applications}

\subsubsection*{Pulse propagation and \textit{ad hoc} formula for saturated absorption}
Due to the range of fluence used in laser induced cavitation, the energy deposition cannot be simply described by a linear Beer-Lambert's absorption law. Instead, we must include saturation effects of the middle infrared (2.9 µm) absorption band by replacing the linear (unsaturated) absorption coefficient $\alpha$ with a fluence dependent one $\alpha^*(F)$. Here $F(\rho,z)=\int I(\rho,z,t')dt'$ is the fluence of a laser beam with a time-dependent intensity $I(\rho,z,t)$ propagating along $z$ and $*$ denotes the saturation. For simplicity we assume the beam is initially Gaussian in space and time such that:
\begin{equation}\label{GaussSTProfile}
I(\rho,0,t)=I_0 e^{-2\frac{\rho^2}{w_0^2}} e^{-2\frac{t^2}{\tau_0^2}}
\end{equation}
where $w_0$ in the beam waist and $\tau_0$ is the duration. In cylindrical coordinates $(\rho,\phi,z)$ the fluence reads $F(\rho,z=0)=F_0\cdot\exp(-2\rho^2/w_0^2)$ where the maximum fluence is  $F_0=I_0\tau_0\sqrt{\pi/2}$.

Even in the presence of saturation, the strong absorption in the 3 µm range results in a very limited penetration depth typically on the order of few central wavelength. In this regime where diffraction phenomena can be neglected, the evolution of $F(\rho,z)$ can be written as
\begin{equation}\label{sat_BL}
\left(\frac{\partial F}{\partial z}\right)_\rho=-\alpha^*(F)F
\end{equation}
As a consequence, Eq.(\ref{sat_BL}) can be solved in the line-of-sight (LoS) approximation, meaning that the fluence profile at any $z$ can be obtained by integrating Eq.(\ref{sat_BL}) as an ODE for a fixed set of $\{(\rho,\phi)\}$ (a given LoS) with the initial condition $F(\{(\rho,\phi)\},z=0)$ corresponding to the incident beam profile. For a Gaussian profile, the $\phi$ dependence can be omitted.

The modulus of the r.h.s in Eq.(\ref{sat_BL}) represents the density of energy deposition (in J$\cdot$m$^{-3}$ or J$\cdot$cm$^{-3}$), i.e. the source term in the transient heat equation. As a consequence, the detailed evolution of $F$ as the beam penetrates into the liquid has a great influence on the (density) energy deposition that is more or less localized and which magnitude can be greatly affected by the form taken by $\alpha^*(F)$.

The fluence dependent absorption coefficient $\alpha^*(F)$ can be written:  
\begin{equation}\label{Sat_Abs_S}
    \alpha^*(F)=\alpha_0+\Delta\alpha \cdot \frac{F/F_s}{1+F/F_s}
\end{equation} 
where $\alpha_0$ is the unsaturated absorption coefficient ($F\rightarrow 0$), $\Delta\alpha$ the drop of absorption ($F\rightarrow \infty$) and $F_s$ the saturation fluence. Eq.(\ref{Sat_Abs_S}) differs from the conventional saturation formula $\alpha^*_{c}(F)=\alpha_0/(1+F/F_s)$.

The conventional saturation law $\alpha^*_{c}(F)$ often used in the literature is based on rate-equations involving few discrete levels and valid only for dilute concentration of absorbing molecules. The \textit{ad hoc} formula (\ref{Sat_Abs_S}) reproduces the simulated characteristic features (sharper transition, non-vanishing residual absorption) when collective effects arise from strong couplings due to high concentration. 

\begin{figure}[htbp]
\centering
\includegraphics[width=1\textwidth]{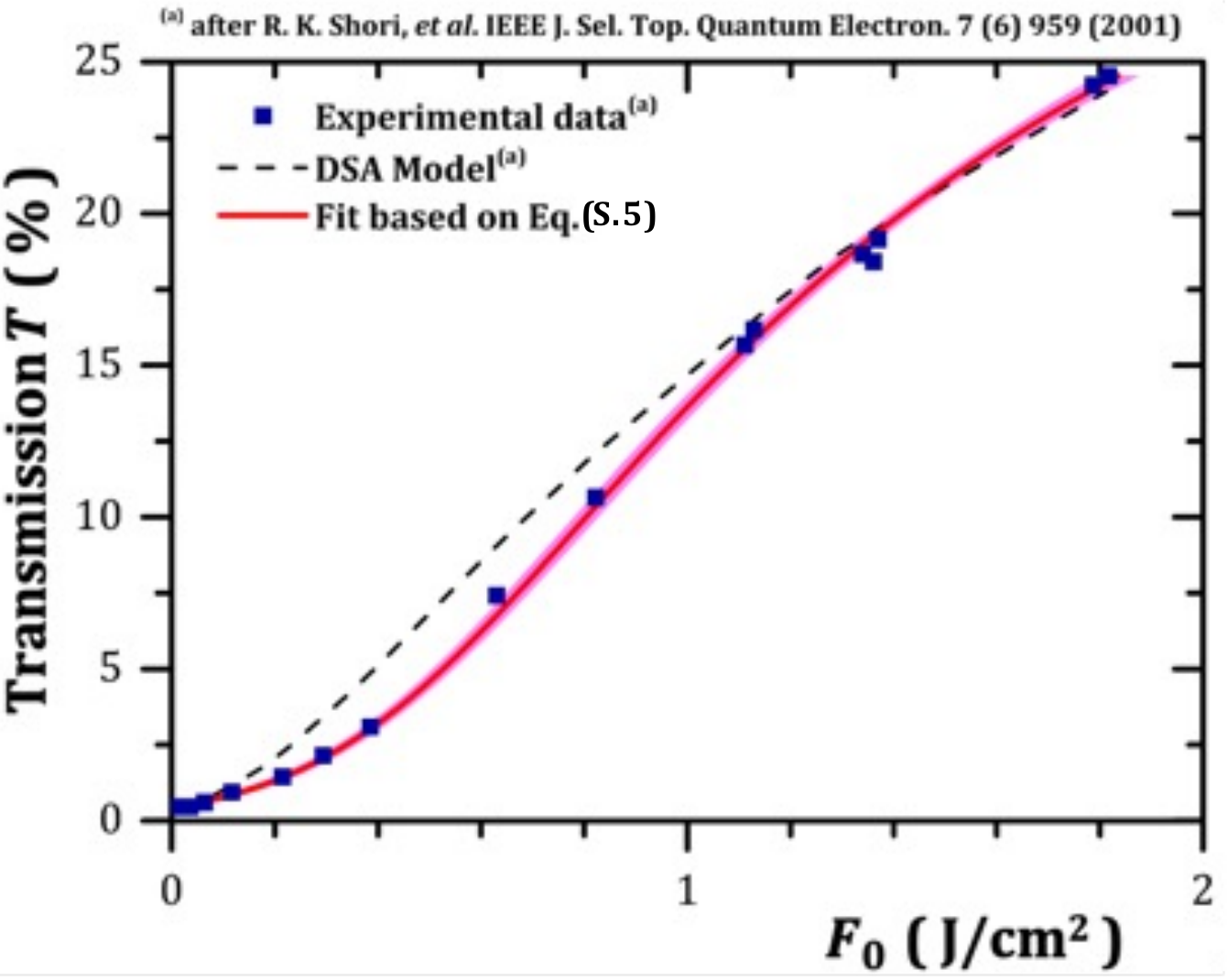}
\caption{\label{FigShoriSat} Saturation of the transmission $T$ (in \%) through a thin water layer (4.26 µm) vs. the fluence $F_0$ of a middle infrared laser (2.94 µm): Experimental measurements (full squares), DSA model prediction (dashed line), Fit (solid red line) within a 95\% confidence band (light red ribbon) based on Eq.(\ref{Shori_T})}
\end{figure}

The Eq.(\ref{sat_BL}) can be integrated using Eq.(\ref{Sat_Abs_S}) leading to:
\begin{equation}\label{Implicit_F_of_z}
    e^{-\alpha_0 z}=\frac{F(z)/F_s}{F_0/F_s}\frac{(1+(1+\eta)F_0/F_s)^\beta}{(1+(1+\eta)F(z)/F_s)^\beta}
\end{equation}
where $\eta=\Delta\alpha/\alpha_0$, and $\beta=\eta/(1+\eta)$. Introducing the fluence $F(L)$ at the output of the sample of thickness $L$ such that $F(L)=T\cdot F_0$ where $T$ is the incident fluence dependent transmission, one can write:
\begin{equation}\label{Shori_T}
F_0=\frac{F_s}{1+\eta}\cdot\frac{1-(T\cdot e^{\alpha_0 L})^{1/\beta}}{(T\cdot e^{\alpha_0 L})^{1/\beta}-T}
\end{equation}
Using Eq.(\ref{Shori_T}), it is thus possible to fit the experimental data (Fig.\ref{FigShoriSat}) obtained by Shori, \textit{et al}. [R. K. Shori, et al. "Quantification
and modeling of the dynamic changes in the absorption coefficient of water
at $\lambda$ = 2.94 µm," IEEE J. Sel. Top. Quantum Electron. 7 (6) 959-970 (2001)]. 
The fit is performed for a fixed thickness $L=$ 4.26 µm. The retrieved fitting parameters are $\alpha_0=12787.3 \pm 18.1 \text{ cm}^{-1}$, $\Delta\alpha= -10906.6 \pm 23.9 \text{ cm}^{-1}$ (and the derived parameters $\eta\simeq-0.85$, and $\beta\simeq -5.78$), and $F_s=0.136 \pm 0.006 \text{ J.cm}^{-2}$.

Although based on the \textit{ad hoc} relationship Eq.(\ref{Sat_Abs_S}), fitting with Eq.(\ref{Shori_T}) results in a much better qualitative and quantitative agreement (see Fig.\ref{FigShoriSat}) with the experiment than the prediction derived from the Dynamical Saturated Absorber (DSA) model developed by Shori, \textit{et al}. This is especially true in the 0-to-1 J/cm$^{-2}$ range of fluence where explosive boiling occurs. Consequently Eq.(\ref{Sat_Abs_S}) provides a much better description for the energy deposition which is key for the bubble formation. As we shall see below, not only Eq.(\ref{Sat_Abs_S}) better reproduces the reported results, but it also allows us to derive analytical expressions leading by thermodynamic considerations to the exact formulation of the excess energy initially available after explosive boiling for bubble expansion and then to the maximum reachable bubble diameter.

\subsubsection*{Effective threshold fluence $F^*_\textrm{eb}$ for explosive boiling}

We first define $F_\textrm{eb}^*$ as the effective critical fluence for which the density of deposited energy $u_\textrm{dep,eb}^*=\alpha^*(F_\textrm{eb}^*)F_\textrm{eb}^*=H_\textrm{eb}\rho_l$ exactly corresponds to the explosive boiling energy density with no excess energy. Considering the expression Eq.(\ref{Sat_Abs_S}) for $\alpha^*(F)$, $F_\textrm{eb}^*$ is given by:
\begin{equation}\label{Effective_Threshold}
    F_\textrm{eb}^*=\frac{1}{2q}\bigg(\sqrt{\Delta F^2+4q F_sF_\textrm{eb}}+\Delta F\bigg)
\end{equation}
where $q=(\alpha_0+\Delta\alpha)/\alpha_0=1+\eta$, $\Delta F=F_\textrm{eb}-F_s$ and $F_\textrm{eb}=H_\textrm{eb}\rho_l/\alpha_0=0.1 \text{ J.cm}^{-2}$ is the threshold in the absence of saturation effect ($u_\textrm{dep,eb}=\alpha_0 F_\textrm{eb}=H_\textrm{eb}\rho_l$). After calculation, we find $F_\textrm{eb}^*=0.21 \pm 0.02 \text{ J.cm}^{-2}$ which is almost twice higher than the threshold in the absence of saturation $F_\textrm{eb}$.

As it propagates, when the local beam fluence $F(\rho,z)>F_\textrm{eb}^*$, the density of deposited energy exceed the specific spinodal energy $H_\textrm{eb}\rho_l$ required to vaporize the water explosively, thus leading the vapour creation. If $F(\rho,z)<F_\textrm{eb}^*$, then the pulse energy is still absorbed resulting in a temperature rise yet remaining below the vaporization limit.

Consequently, there is a region of space $z>0$ defining a volume $V^*$ satisfying $F(\rho,z)>F_\textrm{eb}^*$ where vapour will be created with some excess energy leading to the bubble expansion.  

\subsubsection*{Critical radius $R^*_\textrm{eb}$, penetration depth $z^*_\textrm{eb}$ and the corresponding volume $V^*$ for explosive boiling}

\begin{figure}[htbp]
\centering
\includegraphics[width=1\textwidth]{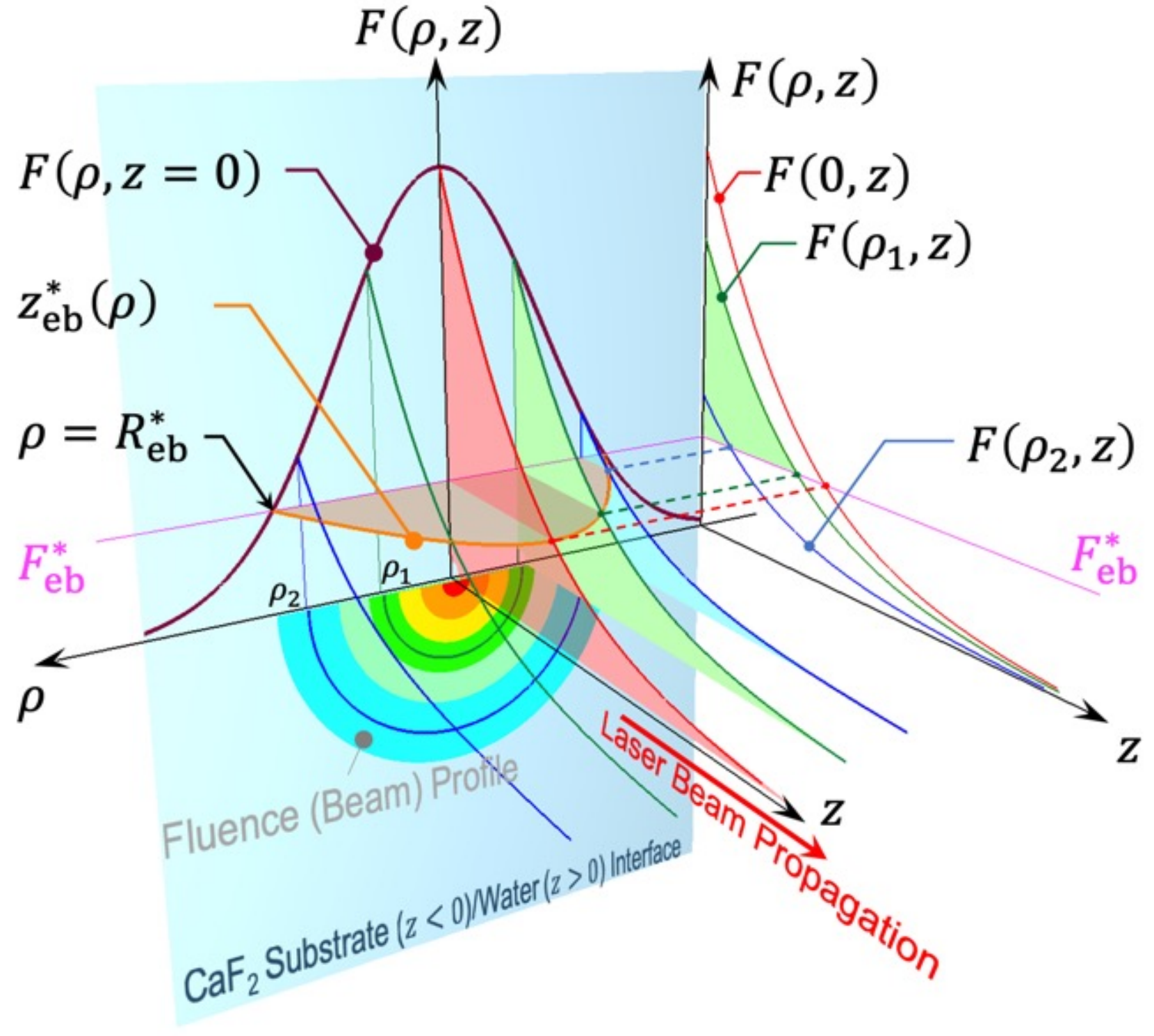}
\caption{\label{FigSVolDef} Evolution of the local fluence $F(\rho,z)$ of Gaussian beam propagating through water ($z>0$) in the presence of saturated absorption. The incident fluence profile $F(\rho,z=0)$ (purple line) is decomposed into several line-of-sight (LoS) $\rho=0$, $\rho_1$, $\rho_2$, $...$ (resp. red, green, blue line). Along each LoS $\rho=\rho_i$, the fluence $F(\rho_i,z)$ evolves according to Eq.(\ref{sat_BL}) with the initial conditions $F(\rho_i,0)$. The solutions $F(\rho_i,z)$ of Eq.(\ref{Implicit_F_of_z}) are projected onto the ($z$,$F$) plane in order to highlight, for each LoS solution, the penetration depth $z=z^*_\textrm{eb}(\rho_i)$ where $F(\rho_i,z)$ is equal to the explosive boiling threshold $F^*_\textrm{eb}$ (magenta line). All the orange shaded region within $0<\rho<R^*_\textrm{eb}$ and $0<z<z^*_\textrm{eb}(\rho)$ (orange line) experience explosive boiling.} 
\end{figure}

In order to evaluate $V^*$ for a Gaussian beam, we first note that at the input $z=0$ where the CaF$_2$-water interface is located (see Fig.\ref{FigSVolDef}), the requirement $F>F_\textrm{eb}^*$ may be written:
$$F(\rho,z=0)=F_0\cdot\exp(-2\rho^2/w_0^2)>F_\textrm{eb}^*$$
Similar to the ablation spot-size formula, we can thus define a critical radius:
\begin{equation}\label{Critical_Radius}
    R^*_\textrm{eb}=w_0\sqrt{\textrm{ln}(F_0/F_\textrm{eb}^*)/2}.
\end{equation}
Within the beam profile, explosive boiling thus can only occur if $\rho<R^*_\textrm{eb}$. 

Secondly, as the beam penetrates into the liquid, its fluence decreases with $z$ (see Fig.\ref{FigSVolDef}). In the LoS approximation, for each $\rho_i$ ($i=1, 2, ...$) defining the LoS considered, the explosive boiling condition $F(\rho=\rho_i,z)>F_\textrm{eb}^*$ is fulfilled up to a certain critical penetration depth $z^*_\textrm{eb}(\rho_i)$ such that $F(\rho_i,z^*_\textrm{eb}(\rho_i))=F_\textrm{eb}^*$ (see projections in Fig.\ref{FigSVolDef}).

We can thus use Eq.(\ref{Implicit_F_of_z}) to evaluate for each LoS (each $\rho$ of constant fluence in the case of an axially symmetric beam) what is the maximum depth $z^*_\textrm{eb}(\rho)$ at which explosive boiling may occur.  This penetration depth (inside the critical radius) for explosive boiling is given by:
\begin{equation}\label{sat_depth}
z_\textrm{eb}^*(\rho<R_\textrm{eb}^*)=\frac{1}{\alpha_0}\textrm{ln}\bigg(\frac{f}{f_\textrm{eb}^*}\frac{(1+(1+\eta)\cdot f_\textrm{eb}^*)^\beta}{(1+(1+\eta)\cdot f)^\beta}\bigg)
\end{equation}
where $f=F(\rho)/F_s$ and $f_\textrm{eb}^*=F_\textrm{eb}^*/F_s$.

\begin{figure}[htbp]
\centering
\includegraphics[width=1\textwidth]{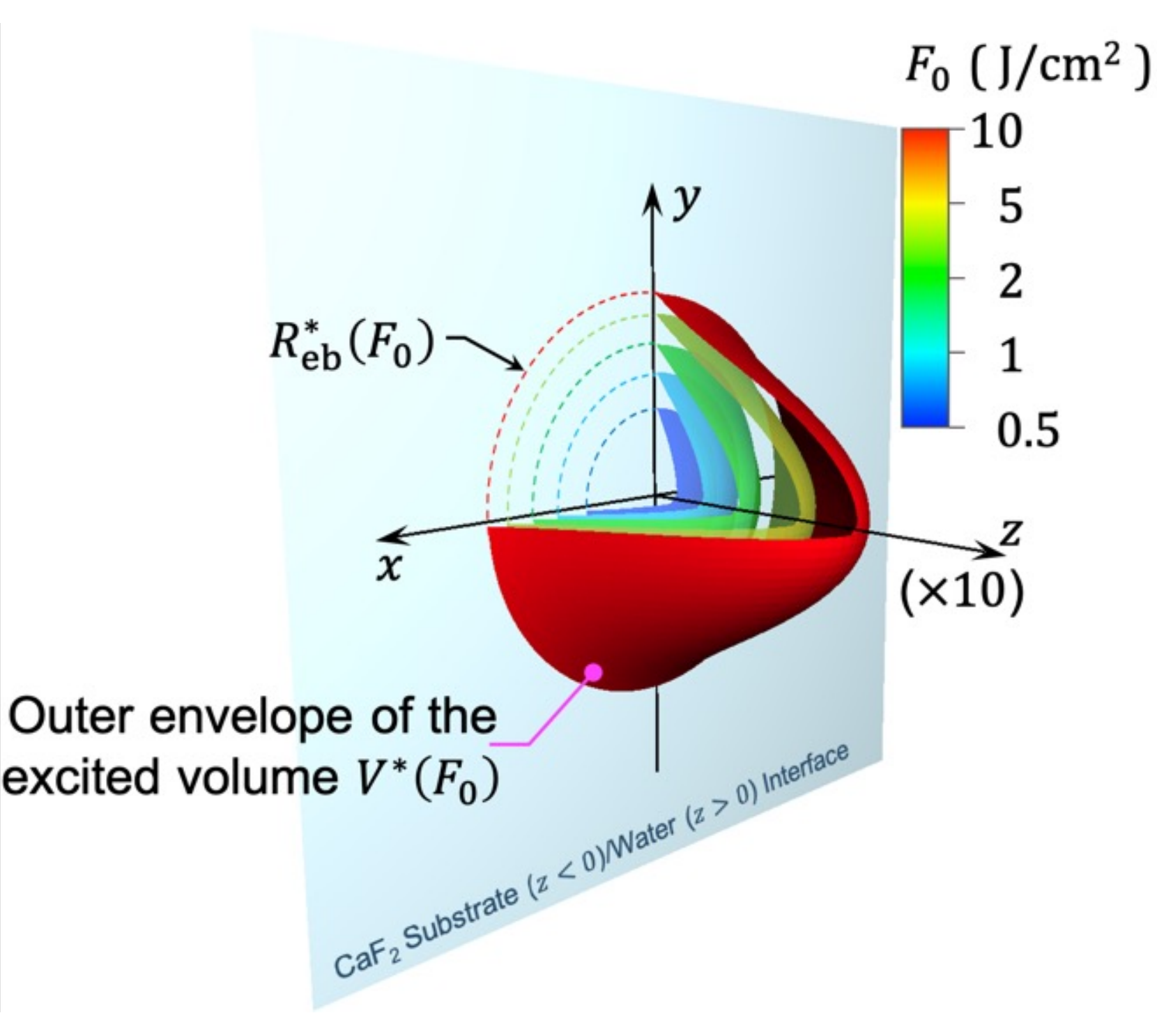}
\caption{\label{FigSVolume_vs_F0} Evolution of the excited volume $V^*$ boundary (outer envelope) as a function of the peak incident fluence $F_0$.
The outer envelopes represented for different $F_0$ being relatively flat, the $z$-axis is stretched (10$\times$). The first octant ($x,y,z>0$) has also been cut away for clarity. The intersection of the envelope with the input interface correspond to the critical radius $F^*_\textrm{eb}$ (dashed lines)}
\end{figure}

Eq.(\ref{sat_depth}) defines a boundary $z_\textrm{eb}^*(\rho)$ for the excited volume $V^*$ (by a $2\pi$ revolution around the $z$ axis of the critical curve $z_\textrm{eb}^*(\rho)$ represented in orange in Fig.\ref{FigSVolDef}). The evolution of this volume envelope as a function of $F_0$ is depicted in Fig.\ref{FigSVolume_vs_F0}.
From a direct integration:
$$V^*=\int_{\rho=0}^{R_\textrm{eb}^*}2\pi z_\textrm{eb}^*(\rho)\rho\textrm{d}\rho$$
we can thus derive the volume $V^*$ where explosive boiling occurs given by:
\begin{equation}\label{sat_volume}
\begin{aligned}
  V^*={} & \frac{\pi w_0^2}{2\alpha_0}\cdot \bigg(\frac{1}{2}\ln^2(F_0/F_\textrm{eb}^*)+\beta \ln\Big(g(F_\textrm{eb}^*)\Big)\cdot\ln(F_0/F_\textrm{eb}^*)\\
  &+\beta\cdot\textrm{dilog}\Big(g(F_0)\Big)-\beta\cdot\textrm{dilog}\Big(g(F_\textrm{eb}^*)\Big)\bigg)  
\end{aligned}
\end{equation} where $g(x)=1+\eta\cdot x/F_s$, and $\textrm{dilog}(x)$ is the dilogarithm function ($\textrm{dilog}'(x)=\ln(x)/(1-x)$).
The variation of $V^*$ as a function of $F_0$ is represented in Fig.\ref{FigSatQty_vs_F0}.(a).

The shape of the volume $V^*$  (Fig.\ref{FigSVolume_vs_F0}) is oblate ($z^*_\textrm{eb}<R^*_\textrm{eb}$) as shown by the aspect ratio $z^*_\textrm{eb}/R^*_\textrm{eb}$ represented as function of $F_0$ in Fig.\ref{FigSatQty_vs_F0}.(b). From this initially flattened volume, vapour creation will lead to hemispherical bubbles seeds that further expand following the Rayleigh-Plesset equation. 

Outside $V^*$, i.e. beyond $z^*_\textrm{eb}$ and outside $R_\textrm{eb}^*$, the remaining pulse energy is still fully deposited (no energy is transmitted for the thickness considered in our experiment) and the water heated up but below any phase change. A fraction of this rapidly diffusing hot outer shell (typically a several tens of micrometer shell thickness from 100 down to 50$^\circ$C) could in principle contribute to water vapour formation in the bubble as the low pressure trailing edge of the shock-wave passes through it. However the energy left in this outer shell represents a smaller and smaller fraction of the total energy deposited as the fluence increases. Consequently, we only expect some discrepancies close to the bubble formation threshold where the process is inherently very sensitive to any fluctuation.

\subsubsection*{Excess energy $E^*_\textrm{ex}$ available for explosive boiling}

The energy $E_\textrm{dep}^*$ deposited within the volume $V^*$ where $F(\rho,z)>F_\textrm{eb}^*$ is the difference between the incident energy $E_\textrm{inc}^*$ that may lead to explosive boiling:
\begin{equation}\label{Input_Energy}
    E_\textrm{inc}^*=\int_{\rho=0}^{R_\textrm{eb}^*}2\pi F(\rho,z=0)\rho\textrm{d}\rho=\frac{\pi}{2}w_0^2\left(F_0-F_\textrm{eb}^*\right),
\end{equation}
and the remain output energy 
\begin{equation}\label{Output_Energy}
    E_\textrm{out}^*=\int_{\rho=0}^{R_\textrm{eb}^*}2\pi F(\rho,z=z^*_\textrm{eb}(\rho))\rho\textrm{d}\rho
\end{equation}
that will further be absorbed but without leading to explosive boiling. Since by definition, $F(\rho,z=z^*_\textrm{eb}(\rho))=F_\textrm{eb}^*$ one can simplify:
\begin{equation}\label{Output_Energy_2}
    E_\textrm{out}^*=\int_{\rho=0}^{R_\textrm{eb}^*}2\pi F_\textrm{eb}^*\rho\textrm{d}\rho=\pi R_\textrm{eb}^{*2}F_\textrm{eb}^*
\end{equation}

A fraction $W_\textrm{eb}^*$ of the deposited energy $E_\textrm{dep}^*$ corresponds to the specific spinodal energy $H_\textrm{eb}\rho_l$ required to vaporize the water explosively, integrated over the volume $V^*$:
\begin{equation}\label{Enthalpy_sat}
    W_\textrm{eb}^*=H_\textrm{eb}\rho_l V^*
\end{equation}
The vapour finally receives an energy excess $E_\textrm{ex}^*(F_0)=E_\textrm{dep}^*-W_\textrm{eb}^*=E_\textrm{inc}^*-E_\textrm{out}^*-W_\textrm{eb}^*$ (Fig.\ref{FigSatQty_vs_F0}.(c)) or in the form given in Eq.(3) of the main text: 
\begin{equation}\label{Excess_Energy_V_sat}
E_\textrm{ex}^*=\int_{\rho=0}^{R_\textrm{eb}^*}2\pi \bigg(F(\rho)-F_\textrm{eb}^*-H_\textrm{eb}\rho_l z_\textrm{eb}^*(\rho)\bigg)\rho\textrm{d}\rho
\end{equation}
The threshold for bubble formation $F_0>F_\textrm{eb}^*=0.2 \text{ J.cm}^{-2}$ is clearly visible on the excess energy $E_\textrm{ex}^*$ as shown in Fig.\ref{FigSatQty_vs_F0}.(c).

As already mentioned, the energy initially available for bubble formation can be slightly higher than $E_\textrm{ex}^*$ since outside $V^*$ exists a layer of heated water (not the point of liquid-vapour spinodal transition) that can boil once the shock wave leaving the excited area induces a low density trailing edge changes the local $(p,T)$ conditions. However since most of the pulse energy is converted into $E_\textrm{ex}^*$ we can neglect this additional fraction.

\subsubsection*{Resulting initial pressure $p^*$}

\begin{figure}[htbp]
\includegraphics[width=1\textwidth]{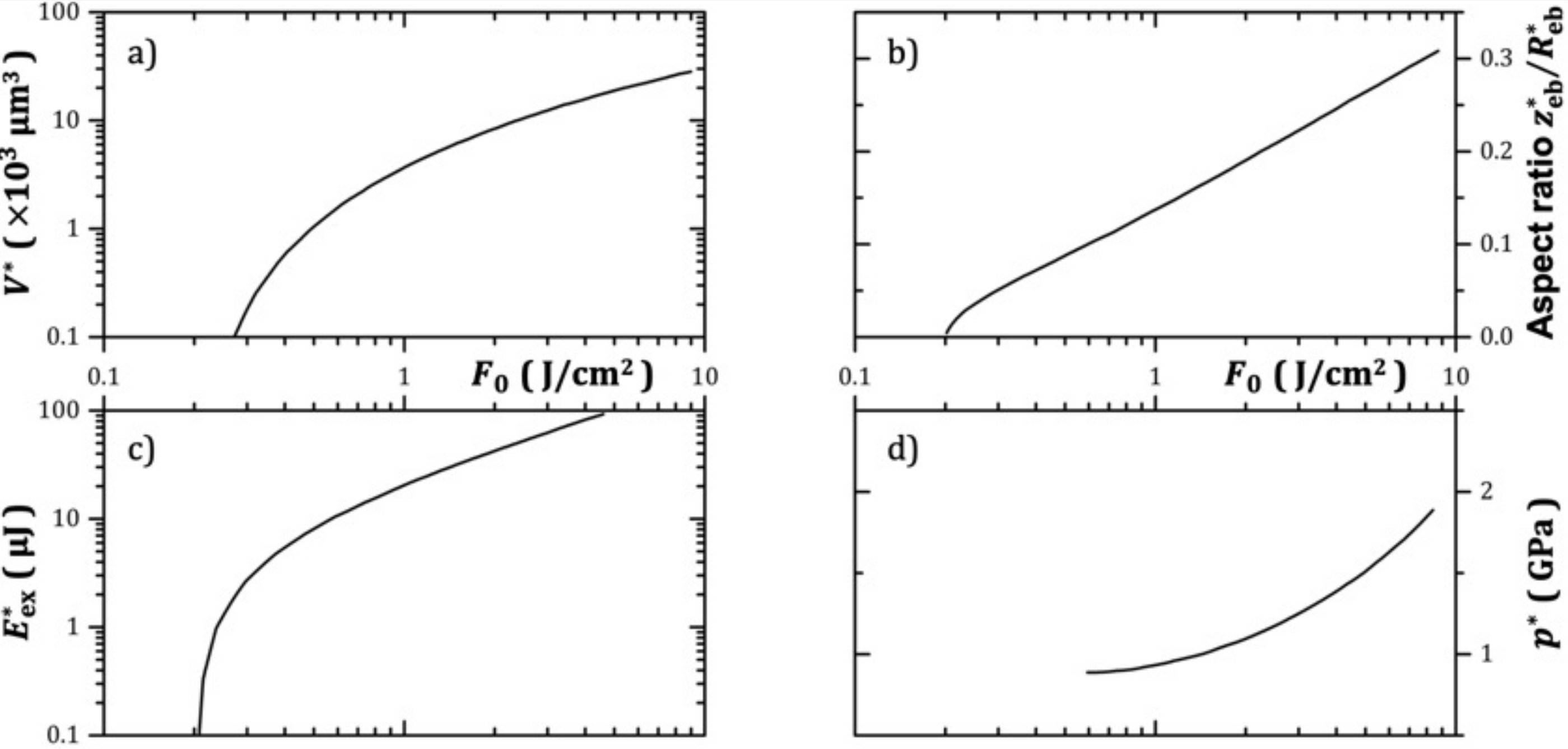}
\caption{\label{FigSatQty_vs_F0} Evolution with the incident fluence $F_0$ of a) the explosive boiling volume $V^*$, b) the aspect ratio $z^*_\textrm{eb}/R^*_\textrm{eb}$ of the volume $V^*$, c) the excess energy $E^*_\textrm{ex}$ contained within $V^*$, and d) the estimated initial pressure $p^*$.}
\end{figure}

The energy $E_\textrm{ex}^*$ (Eq.\ref{Excess_Energy_V_sat}) and volume $V^*$ (Eq.\ref{sat_volume}) allow us to estimate the initial pressure $p^*=E_\textrm{ex}^*/V^*$ at the onset of bubble formation. However such a direct evaluation could be misleading (although correct for estimating an upper limit) since $V^*$ is initially a flat volume while hemispherical bubbles are experimentally observed. A more realistic evaluation of $p^*$ can thus be obtained by substituting $V^*$ by $\Tilde{V}^*=(V^*+2\pi R_\textrm{eb}^{*3}/3)/2$ that accounts for the rapid energy redistribution associated with the doming from the flat excited volume $V^*$ to the hemispherical bubble seed.  Typically here, $p^*$ is in the GPa range (Fig.\ref{FigSatQty_vs_F0}.(d)). This excessively high pressure leads to shock-wave formation thus dissipating some of the deposited energy.

\newpage

\end{document}